# Fourier Optics on Graphene

Ashkan Vakil and Nader Engheta[*]

*Department of Electrical & Systems Engineering*
*University of Pennsylvania*
*Philadelphia, PA 19104, USA*

**Abstract**

Using numerical simulations, here we demonstrate that a single sheet of graphene with properly designed inhomogeneous, nonuniform conductivity distributions can act as a convex lens for focusing and collimating the transverse-magnetic (TM) surface Plasmon polariton (SPP) surface wave propagating along the graphene. Consequently, we show that the graphene can act as a platform capable of obtaining spatial Fourier transform of infra-red (IR) SPP signals. This may lead to rebirth of the field of Fourier Optics on a one-atom-thick structure.

---

[*] To whom correspondence should be addressed. Email: Engheta@ee.upenn.edu

Optical signal processing systems composed of optical elements are often designed to collect, process and transmit information that is generally of a spatial nature [1, 2]. It is well known, for example, that a simple lens, which is the most important component of an optical data processing system, provides a simple and useful way of obtaining spatial Fourier transform. However, the necessity of processing information at high rates and small scales and the ubiquitous ambition to have highly miniaturized optical systems call for novel ideas in this realm. In this Letter, we suggest that graphene—a single atomic layer of carbon atoms [3-7]—can be a new platform for "one-atom-thick" optical signal processing. First, by using graphene as a "flatland" platform, one can shrink the thickness of the optical system down to thickness of a carbon atom. Second, since the optical signals may propagate along the graphene in the form of highly-confined surface-plasmon-polariton (SPP) surface waves with guided wavelength much shorter than the free-space wavelength [8-11], the other two dimensions could essentially be of subwavelength scale, yielding a very compact optical system.

Since the first isolation of individual flakes of graphene in 2004, there has been a massive interest in studying various unique properties of this material [3-5]. Graphene is a very promising material in emerging nanoelectronic applications. However in the present work we consider its interaction with the electromagnetic waves for the purpose of low-dimensional optical signal processing. In order to solve the Maxwell equations in the presence of graphene, we need to utilize an effective parameter that models dynamics of electrons in this material. In describing optical behavior of graphene, one commonly uses complex conductivity, i.e., $\sigma_g = \sigma_{g,r} + i\sigma_{g,i}$, which depends on the radian frequency $\omega$, charged particle scattering rate $\Gamma$ which represents loss mechanism, temperature $T$ and chemical potential $\mu_c$. The Kubo formula is considered here to calculate the conductivity of graphene in abcense of external magnetic field [6-8]. The graphene chemical potential depends on the carrier density, which can be controlled at will by gate voltage, electric or magnetic bias field, and/or chemical doping [6-8].

Thanks to this tunability, one can adjust the complex conductivity of graphene locally. Changing the local level of chemical potential may be managed, in practice, by exploiting several methods, some of which we have proposed in [9]. Depending on the level of chemical potential the imaginary part of conductivity, in particular, obtains different values in different ranges of frequencies [6, 8, 9, 12]. But how could this versatility be beneficial? In [9], we discussed how the proper inhomogeneous conductitivty distribution on a single sheet of graphene may lead to the possibility of taming and manipulating transverse-magnetic (TM) electromagnetic SPP along the graphene. The dispersion relation for such highly confined TM SPP surface waves along a free-standing graphene in free space is derived as [8-11] $\beta^2 = k_0^2 \left[1 - \left(2/\eta_0 \sigma_g\right)^2\right]$, where $k_0$ and $\eta_0$ are, respectively, the free space wavenumber and the intinsic impedance of free space. As can be seen from this dispersion relation, by creating specific distributions of inhomogeneous conductivity, one can have desired patterns of wavenumber (and effective refractive index) for SPP surface waves, offering the possibility of designing IR metamaterials and transformation optical devices on a carbon flatland [9]. Here, as another novel application of graphene, we theoretically demonstrate that by creating a specific pattern of conductivity across a single sheet of graphene, Fourier transform functionality may be achieved in one-atom-thick structure using an IR signal guided as SPP along the graphene.

The notion of designing a one-atom-thick lens on graphene is simply built upon the basic idea of how a regular 3-dimensional optical lens works. When a regular lens is illuminated with a plane wave, as we move away from axis of the lens with higher refractive index than outside medium, phase fronts experience smaller phase difference. This nonuniformity in phase differnce distribution results in curved phase fronts that become focused at focal point of the lens. One can visualize the same phenomenon happening for the SPP surface waves along the single sheet of graphene. Consider, for instance, Fig. 1. As mentioned earlier, to create a nonuniform inhomogeneous conductivity pattern along graphene, three possible methods are suggested in [9]. In Fig. 1, as an example, we have depicted schematically and conceptually one of these proposed methods, i.e., the use of "uneven ground plane" to create an one-atom-thick lentil-shaped inhomogeneity in distribution of the conductivity of a single-layered graphene. In this method, highly doped silicon substrate with uneven height profile may be considered as ground plane. If a fixed DC biasing voltage is applied between the ground plane and graphene, since the distance between the flat graphene and the uneven ground plane is not uniform, the graphene layer will experience different values of local biasing DC electric fields at the sections with different distances from the ground plane, resulting in an inhomogeneous carrier density and in turn inhomogeneous chemical potential distribution which will produce a nonuniform conductivity pattern across graphene layer [13]. In practice the distance between the ground plane and graphene will be filled up with a regular dielectric spacer, e.g. $SiO_2$. Here, however for simplicity in our numerical simulations, we assume that there is no spacer. This lentil-shaped region with different conductivity—and therefore different equivalent SPP refracrive index—can act as a one-atom-thick lens. The shape of the inhomogeneity is chosen similar to cross section of a regular optical double-convex lens, but other shapes can be considered as well. We know that, at its back focal plane, a double-convex lens provides the Fourier transform of a function located at its front focal plane. Here we show that the proposed inhomogeneity on graphene also does so. In order to guarantee that the Fourier transformation property holds for our proposed flat lens, the following conditions must hold: (I) The lens must obtain the Fourier transform of a point-like object—which generates circular phase fronts of the SPP on the graphene—placed at its front focal point as linear phase fronts at exit, i.e., $\mathcal{F}\{t(z) = \delta(z)\}|_{f_z = \frac{k_z}{2\pi}} = 1$, where $\mathcal{F}$ and $t(z)$ denote, respectively, the spatial Fourier transform and transmittance of the object located at the front focal point, and $f_z$ and $k_z$ are spatial frequency and wavenumber, respectively. (II)

The lens must also obtain the Fourier transform of a uniform object—which generates uniform linear phase fronts—placed at the front focal line, as circular phase fronts converging at back focal point of the lens, i.e., $\mathcal{F}\{t(z)=1\}|_{f_z=\frac{k_z}{2\pi}} = \delta(\frac{k_z}{2\pi})$. (III) The lens output must stay invariant, except for a linear phase shift, with respect to shift in the input in the transverse direction; in other words, moving the object along the front focal line must result only in a linear phase variation in the spatial frequency domain at the back focal line. That is: $\mathcal{F}\{t(z)=\delta(z-z_0)\}|_{f_z=\frac{k_z}{2\pi}} = e^{iz_0 k_z}$. Finally, (IV) the lens must provide the Fourier transform of an object with a uniform intensity and a linear phase variation located at the front focal line as a converging circular phase fronts to a point shifted along the back focal line, i.e., $\mathcal{F}\{t(z)=e^{izk_0}\}|_{f_z=\frac{k_z}{2\pi}} = \delta(\frac{k_z}{2\pi}-\frac{k_0}{2\pi})$. Verifying conditions stated above, one can assure that the introduced inhomogeneity on the graphene can indeed perform as a lens. In line with our intuition, our numerical simulations demonstrate that the inhomogeneity indeed provides the above properties, confirming that the one-atom-thick lens provides the Fourier transform of the object at its focal line—since everything here occurs on a monolayer, the 2-dimensional object plane of regular optics collapses to a line, hence the descriptor focal "line".

In the rest of this Letter, we present the results of our numerical simulations using the CST Microwave Studio™ commercial software package [14]. In the simulations, the temperature and the frequency of operation are assumed to be $T = 3$ K and $f = 30$ THz, respectively. We emphasize that we could choose any other temperature—e.g., room temperature—in these simulations and still observe similar qualitative effects. The advantage of choosing $T = 3$ K, however, is that by operating at this temperature the amount of loss in graphene will be much less than the corresponding amount at room temperature. We have also used charged particle scattering rate $\Gamma = 0.43$ meV in our calculations, consistent with [6, 7]. At $T = 3$ K and $f = 30$ THz, we would like to have the chemical potential of the "background" graphene layer to be maintained at $\mu_c$ = 150 meV, corresponding to complex conductivity $\sigma_1 = 0.0009 + i0.07651$ mS (using the Kubo formula [6, 7]). For this conductivity for the wavenumber for the TM SPP given above, we have $\Re(\beta_{SPP,1}) = 69.34 k_0$ and $\Im(\beta_{SPP,1}) = 0.71 k_0$—equivalently effective SPP index, which we define as $\Re(\beta)/k_0$, is $n_{SPP} = 69.34$.

In order to create a desired inhomogeneity in a specific region on graphene to act as a double-convex one-atom-thick lens with higher index for the SPP surface waves, the chemical potential must be changed across that specific region. One may change the local conductivity by using any one of the techniques we proposed in [9] (e.g., having larger separation between the ground plane beneath the desired region (Fig. 1); the separation $d_2$ between the graphene and the ground plane beneath the lens region is larger than separation $d_1$ beneath the background region of the graphene and the ground plane, resulting in a higher index for the lens region compared to "outside", i.e., background graphene region ($n_{SPP,1} < n_{SPP,2}$). For example, if one can attain the chemical potential $\mu_c$ = 120 meV in the lens region, corresponding value of the complex conductivity in that region would be $\sigma_2 = 0.0007 + i0.05271$ mS. For this conductivity, we have $\Re(\beta_{SPP,1}) = 100.61 k_0$ and $\Im(\beta_{SPP,1}) = 0.64 k_0$—equivalently effective SPP index is $n_{SPP} = 100.61$. The lens dimensions for geometries on the graphene used in simulations of Figs. 2 and 3 are as following: $L = 13.3 \lambda_{SPP,1} \approx 1.916\ \mu m$, $L_{lens} = 2.7 \lambda_{SPP,1} = 4 \lambda_{SPP,2} \approx 386.8$ nm, and [15]. $w = 10 \lambda_{SPP,1} \approx 1.444\ \mu m$

Fig. 2 shows the simulation results of a case in which circular SPP waves generated from a point source, guiding along the graphene and impinging onto the proposed double convex lens. Figs. 2a and 2b demonstrate, respectively, the snap shot in time and distribution of the phase of the transverse component of the electric field across the graphene layer. These results clearly demonstrate that the output of the proposed lens is almost a linear SPP wave at the exit of the lens. So the condition (I) is satisfied. By post processing the simulation results, we estimate the focal length of the proposed lens to be around $f = 4\lambda_{SPP,1} \approx 580.2$ nm. Figs. 2c and 2d show the snap shot in time and phase patterns of the transverse component of the electric field for the case that the location of point source is shifted down $2\lambda_{SPP,1}$ on the object line. As it is clear, the effect of shift appears as a linear phase shift in the spatial frequecy domain—on the back focal line. This confirms that condition (III) holds.

Now we consider the scenario in which the lens is illuminated with a guided SPP surface wave with a linear phase front (see Fig. 3). Our proposed lens tranforms such SPP "line" waves into converging circular SPP waves. Figs. 3a and 3b display, respectively, the snap shot in time and phase pattern of the transverse component of the electric field across the graphene sheet. We can observe that at the output of lens, circular SPP waves converge at the focal point of the lens. This is a verification of condition (II). In Figs. 3c and 3d, we present the snap shot in time and distribution of the phase of the transverse component of the electric field for the case of oblique incidence. The effect of shift appears as linear phase shift on the image line as the focus moves along the back focal line, asserting that condition (IV) mentioned above holds.

Finally, we consider a one-atom-thick analog of an archetypal scenario from Fourier optics: a 4f system [1]. A 4f system consists of two identical lenses that are $2f$ apart from each other. This device, which is four focal lengths long, can recover a replica of an object, with even transmittance function of space, placed at one focal length behind the first lens, at one focal lenght in front of the second lens (since $\mathcal{F}\{\mathcal{F}\{t(z)\}\} = t(-z)$). The first lens yields the Fourier transform of the object at its back focal line. In turn, the second lens performs another Fourier transform, delivering a duplicate of the object.

The 4f system is of immense significance for optical signal processing purposes, since it is building block of a `4f correlator', which has important application in implementing the mathematical operations of cross-correlation and convolution [16]. The 4f correlator also serves a wide variety of image processing operations such as spatial filtering of optical signals [1, 2]. In Fig. 4, the simulation results are presented for a one-atom-thick "4f optical system". The first lens is illuminated with guided SPP waves with linear phase front. As can be seen, the illuminated waves are approximately recovered at the exit of the second lens.

In conclusion, our theoretical study indicates that graphene may serve as a low-loss platform for Fourier Optics functionalities in single-atomic-layered structure. To achieve such functionalities, we showed that we must create inhomogeneous nonuniform patterns of conductivity on graphene to control and redirect SPP surface waves as desired. Moreover, since the guided SPP wave is highly confined to the graphene surface, the entire signal processing operation is effectively achieved in extremely thin volume around the graphene. As a result, one can envision several parallel graphene sheets closely packed (but far enough apart not to affect the conductivity dispersion of each single sheet) to have parallel and independent optical signal processing. This unique platform could open new vistas in nanoscale and photonic circuitry and massively parallel platforms for high-speed information processing.

This work is supported in part by the US Office of Scientific Research (AFOSR) Grant numbers FA9550-08-1-0220 and FA9550-10-1-0408.

# Figures

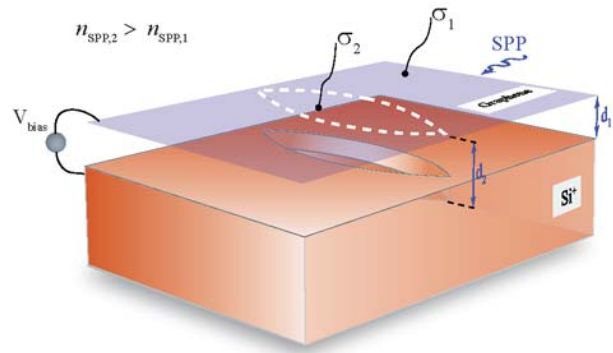

**Fig. 1**. Sketch of our proposed one-atom-thick lensing system for optical Fourier transforming, consisting of a single sheet of graphene with inhomogeneous conductivity distributions. Such inhomogeneity may be achieved by several techniques [9], e.g., by having a highly doped silicon substrate with uneven height profile as ground plane.

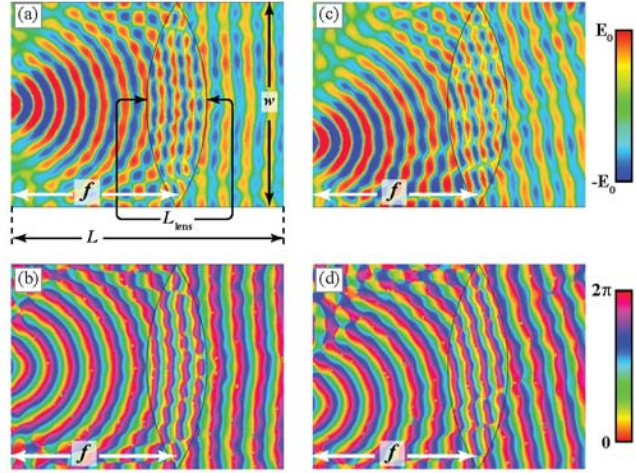

**FIG. 2.** Simulation results for Fourier transforming on graphene. (a) Top view of the snap shot in time of transverse component of the electric field of guided SPP wave for the case of point source illumination, resulting in guided SPP circular waves that propagate through and along the one-atom-thick lens, and exiting as a guided SPP "line" wave. (b) Phase pattern of this guided SPP wave, clearly demonstrating the "point" to "line" Fourier transformation in (a). (c) Top view of the snap shot in time of transverse component of the electric field of the SPP wave for case (a), where the point source is shifted along the left focal line. (d) Phase pattern of the SPP wave in part (c), showing how the shift in the position of the point source in the left focal line results in the phase shift in the SPP line wave in the exit at the right focal line.

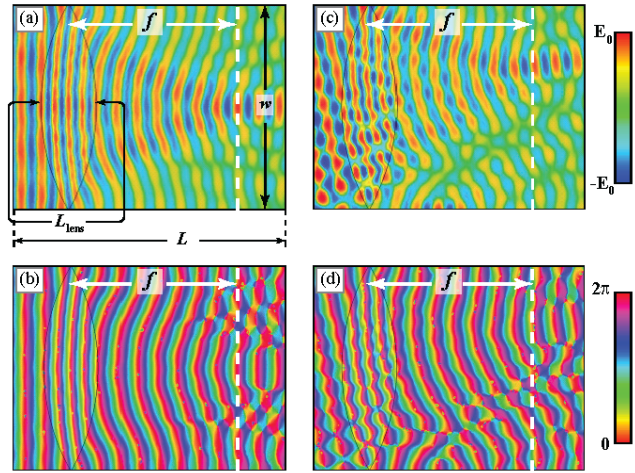

**Fig. 3**. Simulation results for the same graphene lens, but for the incident guided SPP "line" wave: (a) Top view of the snap shot in time of transverse component of the electric field of the SPP line wave incident on lens from the left, resulting into a guided SPP circular wave converging into the focal point on the right. (b) Phase pattern of this SPP wave in (a). (c) Top view of the snap shot in time of transverse component of the electric field of SPP "line" wave obliquely incident on the left from left. (d) Phase pattern of this SPP wave in (c), demonstrating the "line" to "point" Fourier transformation and how the phase shift in the SPP line wave on the left can translate into the shift in the location of the focal point on the right.

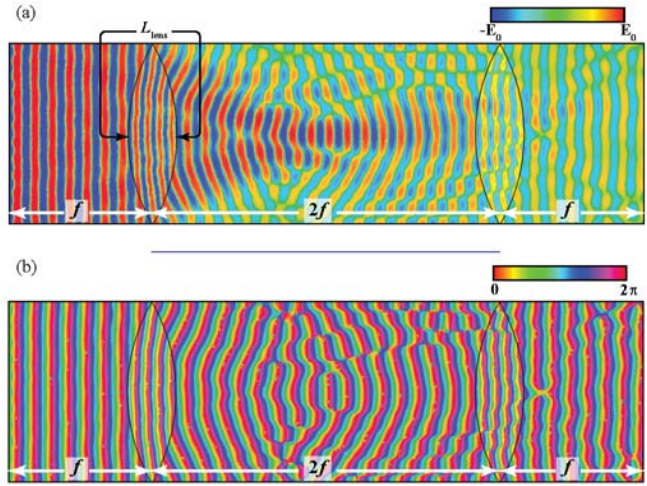

**Fig. 4**. A 4f System on graphene: (a) Top view of the snap shot in time of transverse component of the electric field for a guided SPP "line" wave incidence on the system from left. (b) Phase pattern of this SPP wave propagating through the 4f system, clearly showing how this one-atom-thick 4f system transforms the guided SPP wave from "line" to "point, and then to "line" again.